\documentstyle[aps,preprint]{revtex}

\title{Selfdual backgrounds in $N=2$ five-dimensional Chern-Simons
Supergravity}

\author{M\'aximo Ba\~nados\thanks{Permanent address: Departamento de
F\'{\i}sica, P. Universidad Cat\'olica de Chile, Casilla 306, Santiago 22,
Chile.}}
\address{Departamento de F\'{\i}sica, Universidad de Santiago de
Chile, Casilla 307, Santiago 2, Chile \\ {\tt mbanados@maxwell.fis.puc.cl}}

\begin{document}
\maketitle

\begin{abstract}
We consider five-dimensional $S(2,2|N)$ Chern-Simons supergravity on $M_4\times
S_1$. By fine-tuning the Kaluza-Klein reduction to make the four-dimensional
cosmological constant equal zero, it is shown that selfdual curvatures on
$M_4$ provide exact solutions to the equations of motion if $N=2$.

\

\end{abstract}

At first sight, the 3-dimensional Chern-Simons field theory defined by
the equations of motion
\begin{equation}
g_{ab} F^b=0
\label{3d}
\end{equation}
seems to be completely trivial. However, it is now well-known that
these equations and their generalizations including Wilson lines
contain a large amount of information. Most notably, Witten
\cite{Witten89} has shown that all known knots invariants can be
described as expectation values of the corresponding Wilson lines.
Another application of (\ref{3d}) is three-dimensional gravity. Choosing
the Lie algebra to be $SO(2,2)$, Eqns. (\ref{3d}) are
equivalent to Einstein equations in three dimensions.
[In (\ref{3d}), $F^a$ represent the Yang-Mills curvature 2-form for a
Lie algebra $G$, and $g_{ab}$ is a $G-$invariant tensor.]

Equations (\ref{3d}) have a known generalization to five dimensions,
\begin{equation}
g_{abc} F^b \mbox{\tiny $\wedge$} F^c = 0,
\label{5d}
\end{equation}
where $g_{abc}$ is a $G-$invariant rank three tensor. Contrary to the
three-dimensional case, these equations do not seem to have such a
wide class of applications. In fact the structure of the space of
solutions is largely unknown, and only some special aspects have been
studied.

One of the aspects of (\ref{5d}) that has been studied in some detail is its
interpretation as a five-dimensional gravitational theory
\cite{Chamseddine,BTZ2,TZ,Sardinia,Mora}. If the  Lie algebra is taken to be
$SO(4,2)$, this theory has a sensible interpretation as a gravitational theory
with a Gauss-Bonnet coupling \cite{Chamseddine}. Extending $SO(4,2)$ to
$SU(2,2|N)$ \cite{Nahm} we arrive at a supergravity theory in five dimensions
\cite{Chamseddine}.  Contrary to the three- dimensional case, the relation
between the $d=5$ Chern-Simons and standard supergravities is not
known (see \cite{Sardinia} for a recent discussion). The interesting feature of
Chern- Simons supergravity is that the expected supergroups are incorporated
from the very beginning, $SU(2,2|N)$ and $OSp(32|1)$ in five and eleven
dimensions respectively, and the action is invariant by construction\cite{TZ}.
The eleven dimensional case has been conjectured to be related to M-Theory in
\cite{TZ,Horava}.

The main difference between (\ref{3d}) and (\ref{5d}) is that the
latter has non-flat solutions. Moreover, propagating degrees of
freedom can be shown to exist \cite{BGH}.  A ``natural" background to
study (\ref{5d}) is the flat solution $F^a=0$. However, this phase is
degenerate because small perturbations around $F^a=0$ are completely
trivial.  In fact, the linearized equations  become an identity
\footnote{An interesting corollary of this fact is that the pp-wave
solutions studied in \cite{Boulware-D}, satisfying a linear wave equation,
become trivial at the Chern-Simons point. In other words, when the
coefficients are chosen as in the Chern-Simons action \cite{Chamseddine,BTZ2,TZ}
the wave equation is lost.}. Even though a generic Hamiltonian analysis does exhibit
the presence of local excitations \cite{BGH}, no propagating solutions have been
found so far.

In this paper we study the $SU(2,2|N)$ Chern-Simons theory on
$M_4 \times \Re$ and point out that selfdual curvatures on $M_4$ provide exact
solutions to the equations of motion if $N=2$.

We start by recalling the construction of the $SU(2,2|N)$ Chern-Simons theory.
The $SU(2,2|N)$ 1-form gauge field $\Omega=\Omega_\mu dx^\mu$  is a graded
$(N+4)\times (N+4)$
complex matrix satisfying
\begin{equation}
(\Omega \Gamma )^\dagger = - \Omega \Gamma, \ \ \ \ \ \
\mbox{STr} (\Omega)=0
\label{su}
\end{equation}
where
\begin{equation}
\Gamma = \left( \begin{array}{cc}   \gamma_0 &  0  \\
                           0 &  \mbox{I}_N   \end{array} \right).
\label{Gamma}
\end{equation}
I$_N$ is the identity operator in $N$ dimensions and $\gamma_0 =
\mbox{diag}(1,1-1,-1)$ \footnote{We use the Dirac matrices in five
dimensions $\{\gamma_A,\gamma_B\}=-2\eta_{AB}$ with $\eta_{AB} =
\mbox{diag}(-1,1,1,1,1)$ and $\gamma_5=-\gamma_0\gamma_1\gamma_2\gamma_3$.}.
The maximum number of supersymmetries is $N=4$
(${\cal N}=8$) \cite{Nahm}. Conditions (\ref{su}) can be implemented by
expanding $\Omega$ in the form \cite{Chamseddine}
\begin{equation}
\Omega = \left( \begin{array}{cc}   W &   \psi   \\
                          -\bar\psi &  {\cal A}   \end{array} \right)
\label{Omega}
\end{equation}
where $W\in U(2,2)$ i.e., $(W\gamma_0)^\dagger = -W\gamma_0$, $\psi$ is a Dirac
spinor in five dimensions ($\bar \psi = \psi^\dagger \gamma_0$) and ${\cal A}
^\dagger = - {\cal A}$ is a $N\times N$ matrix.  Using Dirac matrices,
$W$ and ${\cal A}$ can be further expanded as,
\begin{eqnarray}
W &=& {i \over 4} A \, \mbox{I}_4 + {i \over 2} e^A \gamma_A - {1 \over 4}
w^{AB} \gamma_{AB}, \label{W} \\
{\cal A} &=& {i \over N} A\, \mbox{I}_N + {\cal A'},
\end{eqnarray}
with $ {\cal A'} \in SU(N)$, i.e., Tr${\cal A'}=0$. It is direct to check that
conditions (\ref{su}) are indeed satisfied.

The 1-form $e^A$ is identified with the five-dimensional
vielbein\footnote{$e^A$ appears divided by the AdS
radius which we set equal to one in order to simplify the notation.},
$w^{AB}$ is the spin connection and $A$ an Abelian $U(1)$ field. The Lagrangian
has the usual five-dimensional Chern-Simons form. The explicit expression can
be found in \cite{Chamseddine,TZ}.

Our main concern will be the equations of motion which, after setting
$\psi=\bar\psi={\cal A'}=0$, take the form (see \cite{mbanados} for a detailed
derivation of these equations),
\begin{eqnarray}
\epsilon_{ABCDE} \bar R^{AB} \mbox{\tiny $\wedge$} T^C &=& \bar R_{DE} \mbox{\tiny
$\wedge$} F
\label{2} \\
\epsilon_{ABCDE} \bar R^{AB} \mbox{\tiny $\wedge$} \bar R^{CD} &=& - 4  T_E
\mbox{\tiny $\wedge$} F
\label{1} \\
{1 \over 2} \bar R^{AB} \mbox{\tiny $\wedge$} \bar R_{AB} - T^A \mbox{\tiny
$\wedge$}
T_A &=& 4 (N^{-2}-4^{-2}) F \mbox{\tiny $\wedge$} F \label{3}
\end{eqnarray}
where $T^A = De^A$ is the torsion tensor, $F=dA$, and
\begin{equation}
\bar R^{AB} = R^{AB} + e^A \mbox{\tiny $\wedge$} e^B
\label{Rbar}
\end{equation}
with $R^{AB}=dw^{AB} + w^A_{\ C}w^{CB}$.  Note that all coefficients in
(\ref{2},\ref{1},\ref{3}) are fixed (up to trivial rescallings) by
supersymmetry.

The anti-de Sitter (AdS) background has zero torsion, $T^A=0$, and
constant spacetime curvature, $R^{AB} + e^A\mbox{\tiny $\wedge$} e^B=0$. This
configuration
solves
the equations if, and only if,
\begin{equation}
(N-4) F \mbox{\tiny $\wedge$} F=0.
\end{equation}
Anti-de Sitter space is then a solution only for the theory with the maximum
number of supersymmetries $N=4$. The solution with $F\mbox{\tiny $\wedge$} F=0$ is
discarded
because it corresponds to a degenerate background \cite{BGH}. On the contrary,
the phase for which  $F$ has maximum rank, which is allowed by the $N=4$
equations of motion, carries the maximum number of degrees of freedom
\cite{BGH}.

The $N=4 $ theory is interesting in many other respects. For example
under appropriated boundary conditions it can be shown that the group
of asymptotic symmetries is the $WZW_4$ algebra
\cite{Nair-,BGH,Losev,Gegenberg-K}. This
yields  a natural extension to five dimensions of the known relation
between 3d Chern-Simons theory and the affine Kac-Moody algebra.  In
the context of the AdS/CFT correspondence, the $N=4$ theory seems to
contain the correct ingredients; an AdS background and the maximum
number of supercharges.  Note also that $SU(2,2|4)$ represents both
the super ${\cal N}=8$ AdS group in five dimensions and the
superconformal ${\cal N}=4$ group in four dimensions (see
\cite{Gunaydin} and references therein for a detailed study of this group).

Despite the interesting properties of the AdS solution and its
associated supergravity theory, it is certainly not the only solution
to the equations of motion. In order to further analyze the structure
of Chern-Simons gravity we have studied other backgrounds with
different number of supersymmetries. In particular, we consider the Chern-
Simons theory on $M_4\times \Re$.

We shall perform a Kaluza-Klein reduction of equations (\ref{2},\ref{1},\ref{3}).
Our procedure will differ from standard Kaluza-Klein theory in that we do not impose
the torsion condition. Instead, we shall dimensionally reduce the metric and
connection separately.  (The standard, second order, Kaluza-Klein
reduction of 5d gravity with a Gauss-Bonnet term has been studied, for example, in
\cite{KK} and leads to a non-linear electromagnetism.)

The reduction of spacetime indices is trivial.  Let $\Omega_\mu dx^\mu=\Omega_i dx^i
+ \Omega_5 dx^5$.  Equations (\ref{2},\ref{1},\ref{3}) are 4-forms in 5 dimensions
with the structure $\epsilon^{\mu\nu\lambda\rho\sigma} {\cal E}_{\mu\nu\lambda\rho}=
0$. The projection to $M_4$ is achieved by setting the free index $\sigma$ equal to
5 obtaining $\epsilon^{ijkl} {\cal E}_{ijkl}=0$. The remaining equations ($\sigma=
i$) can be solved by the gauge condition $\Omega_5=0$ plus the ``zero mode"
condition $\partial _5\Omega_i=0$ (see \cite{BGH} for more details on the 4+1
decomposition of 5d Chern-Simons theory). In the following, we regard
(\ref{2},\ref{1},\ref{3}) as 4-forms in 4 dimensions.

The tangent indices are split in a similar way, $A= \{a,5\}$ with $a= \{0,1,2,3\}$.
The vielbein has components $\{e^a,e^5\}$ and the spin connection
$\{w^{ab},w^{a5}\}$.  The pair $e^a$ and $w^{ab}$ are naturally the four dimensional
gravitational variables. The 1-form  $e^5$ will be related by the equations of
motion to the gauge field $A$, just as in Kaluza-Klein theory.  The 1-form
$w^{a5}$, on the other hand, has a less clear role.  Note that both $e^a$ and
$w^{a5}$ transform as vectors under $SO(3,1)$ Lorentz rotations acting on the
index $a$. We then led to consider the ansatz,
\begin{equation}
w^{a5}= \sigma_1 \, e^a
\label{w=e}
\end{equation}
where $\sigma_1$ is a constant with dimensions length$^{-1}$.  (Since we have
set the AdS radius equal to one, $\sigma_1$ is actually dimensionless.)

The parameter $\sigma_1$ can be fine-tuned in order to neutralize the
four-dimensional cosmological constant. In fact, the cosmological constant enter in
the equations only through the combination (\ref{Rbar}). The four dimensional
components of the curvature tensor are $\bar R^{ab}= dw^{ab} + w^a_{\ C} w^{C
b} + e^a e^b$. Using (\ref{w=e})  we find,
\begin{equation}
\bar R^{ab} = R^{ab} + (-\sigma_1^2+1)\, e^a \mbox{\tiny $\wedge$} e^b
\end{equation}
where $R^{ab} = dw^{ab}+w^a_{\ c}w^{cb}$ is the intrinsic 4d curvature
2-form.  Thus, if  $\sigma_1=\pm 1$  there is no cosmological constant on
$M_4$.

Two useful formulae that follows from (\ref{w=e}) are
\begin{equation}
T^5= de^5, \ \ \ \ \ \ \ \ \sigma_1 \bar R^{a5}  =\bar T^a,
\label{ident}
\end{equation}
where $\bar T^a= De^a + \sigma_1 e^a \mbox{\tiny $\wedge$} e^5$ denotes $T^A$ with
$A=a$.

Inserting this ansatz into the equations of motion we find the
reduced set of equations for $e^a$, $w^{ab}$ and $A$,
\begin{eqnarray}
 \epsilon_{abcd} R^{ab}\mbox{\tiny $\wedge$} T^5 &=&R_{cd}\mbox{\tiny $\wedge$} F
 \label{1'} \\
\epsilon_{abcd} R^{ab}\mbox{\tiny $\wedge$}  R^{cd} &=& -4 T_5
\mbox{\tiny $\wedge$} F \label{2'} \\
R^{ab}\mbox{\tiny $\wedge$} R_{ab} - 2 T^5\mbox{\tiny $\wedge$} T_5 &=&
8(N^{-2}-4^{-2}) F\mbox{\tiny $\wedge$} F
\label{3'}\\
 \epsilon_{abcd} R^{ab}\mbox{\tiny $\wedge$} \bar T^c &=&
 \sigma_1\, \bar T_d\mbox{\tiny $\wedge$} F
\label{4'}.
\end{eqnarray}
The configuration $R^{ab}=F=T^5=0$ is a solution to these
equations reflecting the absence of a 4d cosmological constant. But we are
interested in other backgrounds for which $F$ has the maximum rank.

By analogy with Kaluza-Klein theory and motivated by (\ref{1'}) we consider
solutions on which the 1-form $e^5$ is proportional to the gauge field $A$. In
this case, the curvature $R^{ab}$ will be selfdual. Real selfdual tensors exist
only on Euclidean space and thus it would be more natural to analytically
continue the above equations to Euclidean signature. However, defining super
AdS in Euclidean space is not
easy. Since we are interested in the full supergravity theory, in particular
the existence of Killing spinors, we will work with Minkowskian signature and
consider complex solutions to the equations of motion. We stress that from the
purely bosonic point of view most imaginary units `$i$' appearing below can be
avoided in the Euclidean sector.

A solution to (\ref{1'}) is given by
\begin{equation}
 T^5 = {i \sigma_2 \over 2} F, \ \ \ \ \  { i\sigma_2 \over 2}
 \epsilon_{abcd} R^{ab} = R_{cd}
\label{T=F}
\end{equation}
where $\sigma_2$ takes the values $\pm 1$, and determines whether $R^{ab}$ is
selfdual or antiselfdual.

Replacing (\ref{T=F}) into (\ref{2'}) we find the condition on the
gravitational and gauge anomalies,
\begin{equation}
R^{ab} \mbox{\tiny $\wedge$} R_{ab} = F\mbox{\tiny $\wedge$} F.
\label{RR=FF}
\end{equation}
Next, we replace all together in (\ref{3'}), and discarding a non-zero factor
of $N$, we find
\begin{equation}
(N-2)\, F\mbox{\tiny $\wedge$} F = 0,
\label{N=2}
\end{equation}
which becomes an identity at $N=2$. Certainly this condition on the
number of extended supersymmetries for the selfdual solutions to exist
deserves a deeper explanation. It is interesting and gratifying that
$N$ is an integer and lies in the range $1\leq N\leq 4$. An independent
derivation for this condition will arise below in the analysis of Killing spinors.

Finally we have to address Eq. (\ref{4'}).  Let $S_a = \epsilon_{abcd} R^{ab}
\mbox{\tiny $\wedge$}
e^c - \sigma_1 F \mbox{\tiny $\wedge$} e_d$. In terms of $S^a$,
Eq. (\ref{4'}) reads $DS^a - (i\sigma_1 \sigma_2/ 2) A\, S^a =0$.
Since the first term in $S_a$ is the Einstein tensor, this equation can be
regarded as a ``generalized" conservation equation for $S_a$. Perhaps the most
interesting solutions are the ``vacuum" ones with $S_a=0$ which implies,
\begin{equation}
 \epsilon_{abcd} R^{ab} \mbox{\tiny $\wedge$}e^c = \sigma_1 F \mbox{\tiny $\wedge$}
 e_d.
\label{Einstein}
\end{equation}

Sumarising, the four-dimensional fields are $e^a,w^{ab}$ and $A$.   The spin
connection is constrained to be selfdual while $A$ is constrained by (\ref{RR=FF}).
The tetrad $e^a$ is determined  by (\ref{4'}) or, in the vacuum case, by
(\ref{Einstein}).

A deep analysis of these equations is beyond the scope of this work, we shall come
back to this point elsewhere \cite{mbanados}. Here we shall exhibit some solutions
in order to prove that they are not self-contradictory.

First, let us point out that the ``unbroken" phase $e^a=0$ solve the above
equations.  Let $w^{ab}$ be any selfdual connection, and $A$ a gauge field such that
(\ref{RR=FF}) is satisfied. The only remaining equation is (\ref{Einstein}) which is
trivially solved by $e^a=0$.

It is natural to ask whether the Eguchi-Hanson \cite{Eguchi-H}, and other selfdual
metrics provide solutions to the above equations. This turns out to be true although
they need to be generalized. The reason is that the zero torsion condition $De^a=
0$ imply $F A= 0$ and then $F\mbox{\tiny $\wedge$}F=0$ (this
follows directly from (\ref{4'})). Since for the Eguchi-Hanson spacetime $R^{ab}
R_{ab}\neq 0$, we conclude that this background will not solve (\ref{RR=FF}).

There is, however, a direct modification to the Eguchi-Hanson metric which satisfies
the full set of equations. Here we shall only present the general argument, for more
details see \cite{mbanados}. Let $e^a_{\mbox{\tiny{EH}}}$ the Eguchi-Hanson tetrad,
and $w^{ab}$ its associated selfdual spin connection computed via
$De^a_{\mbox{\tiny{EH}}}=0$. Let $A$ be a
gauge field such that (\ref{RR=FF}) is satisfied.
A solution to the remaining equation (\ref{Einstein}) can be found by perturbing the
tetrad as $e^a=e^a_{\mbox{\tiny{EH}}}+f^a$. Replacing in (\ref{Einstein}) we obtain,
\begin{equation}
(\epsilon^a_{\ bcd} R^{cd}  + \sigma_1  \delta^a_b \, F) \mbox{\tiny $\wedge$} f^b=-
\sigma_1 F \mbox{\tiny $\wedge$} e^a_{\mbox{\tiny{EH}}}.
\label{f}
\end{equation}
This is an algebraic equation of the form $U_{ab}^{\mu\nu} \, f^b_\nu=
V^\mu_a$. It can be checked \cite{mbanados} that det $U\neq 0$ and thus the
perturbation $f$ is fully determined. Note that the torsion is not zero because
$T^a=De^a=D(e^a_{\mbox{\tiny{EH}}} +f^a)=Df^a\neq0$. For more details on this
construction see
\cite{mbanados}.

A more explicit, although less interesting, family of solutions can
be found as follows. Let $w^{ab} = \Lambda^{ab} B$ where
$B$ is an Abelian 1- form, and $\Lambda^{ab}$ a constant selfdual matrix (note the
position of the indices)
\begin{equation}
\Lambda^a_{\ b} = {1 \over 2}
\left( \begin{array}{cccc}  0   &  1  & 0  & 0 \\
                             1  &  0  & 0  & 0 \\
                             0  &  0  & 0  & i \\
                             0  &  0  & -i & 0
    \end{array} \right).
\label{sol1}
\end{equation}
The curvature becomes $R^{ab} = \Lambda^{ab} dB$ and it is selfdual because
$\Lambda^{ab}$ is. We have chosen in this example $\sigma_2=1$.

Note that since the curvature is selfdual, Eq. (\ref{4'}) can be arranged as a
form-valued eigenvalue equation,
\begin{equation}
 R^a_{\ b}\mbox{\tiny $\wedge$} \bar T^b = -{i \sigma_1\sigma_2
 \over 2} F \mbox{\tiny $\wedge$} \bar T^a
\label{eigenvalue}
\end{equation}
A solution to (\ref{eigenvalue}) is given by,
\begin{equation}
dB= - i F,  \ \ \ \ \ \Lambda^a_{\ b} \bar T^b =
{\sigma_1 \over 2} \bar  T^a
\label{sol2}
\end{equation}
The first relation imply $B=-iA$, while the second
\begin{equation}
d(e^0-\sigma_1 e^1)=0,  \ \ \ \ \  d(i e^2 + \sigma_1 e^3)=0.
\label{sol3}
\end{equation}
The local solution to (\ref{sol3}) depends on two functions $f$ and $g$, $e^0 -
\sigma_1 e^1 = df$ and $ie^2 +\sigma_1 e^3= dg$. Note, in particular, that the
flat space configuration $e^a=dx^a$ is a solution. Finally, we note that
(\ref{RR=FF}) is automatically satisfied,
\begin{equation}
R^{ab}\mbox{\tiny $\wedge$} R_{ab} = -R^a_{\ b} \mbox{\tiny $\wedge$} R^b_{\ a} =
-\mbox{Tr}(\Lambda^2) dB\mbox{\tiny $\wedge$} dB =
F\mbox{\tiny $\wedge$} F .
\end{equation}

Let us now analyze the spinor Killing equation on the selfdual background.
The supersymmetry transformations in this theory are of the form $\delta \Omega
= - \nabla_\Omega \Sigma$ where $\Sigma \in SU(2,2|N)$ and $\Omega$ is given in
(\ref{Omega}). These transformations form a closed algebra and differ from the
standard $d=5$ supergravity formula. The Killing spinor equation is
\begin{equation}
\delta\psi_k = \nabla \epsilon_k - {i \over N} A \epsilon_k = 0, \ \ \ \ \ k=
1,2,...,N
\label{kill}
\end{equation}
where $\nabla = d + W$ is the $U(2,2)$ covariant derivative ($W$ is given in
(\ref{W})). The selfdual solution requires $N=2$. This condition can actually
be re-obtained from the Killing equation.

Applying the derivative $\nabla$ we find the integrability condition,
\begin{equation}
\left({i \over 4} F + {i \over 2} T^A \gamma_A - {1 \over 4} \bar
R^{AB} \gamma_{AB} \right) \epsilon_k = {i \over N} F \epsilon_k
\label{int}
\end{equation}
where the term in brackets is the $U(2,2)$ curvature. The term proportional to
the torsion $\bar T^a$ introduces big complications in this equation because it
depends on the potential $A$ and not only on the curvatures. Fortunately,
thanks to (\ref{ident}) this term can be eliminated by imposing the chirality
condition,
\begin{equation}
\gamma_5\epsilon_k = i \sigma_1 \epsilon_k .
\label{chiral}
\end{equation}
The integrability condition (\ref{int}) has now two sectors depending on the
relative signs of $\sigma_1$ and $\sigma_2$ ($N=2$)
\begin{equation}
-{1 \over 2} R^{ab}\gamma_{ab}\epsilon_k = \left\{
   \begin{array}{cl}      i F \epsilon_k \ \ & \ \ \sigma_1 = \sigma_2, \\
                              0            \ \ & \ \ \sigma_1 = -\sigma_2,
                      \end{array}  \right.
\label{int'}
\end{equation}

On the exact solution (\ref{sol1},\ref{sol2}) with $\sigma_2=1$ this equations
becomes
\begin{equation}
{1 \over 2}(\gamma_0\gamma_1 + i \gamma_2 \gamma_3 ) \epsilon_k =  \left\{
 \begin{array}{cl}     \epsilon_k \ \ & \ \ \sigma_1 = 1, \\
                           0            \ \ & \ \ \sigma_1 = -1,
                      \end{array}  \right.
\label{int''}
\end{equation}
The eigenvalues of $(\gamma_0\gamma_1 + i \gamma_2\gamma_3)/2$ are
$\{1,-1,0,0\}$. Therefore, the sector $\sigma_1=-1$ has two solutions, and it
can be proved that both of them satisfies the chirality condition
(\ref{chiral}). This sector preserves 1/2 supersymmetries.  On the other hand,
the sector $\sigma_1= 1$ has only one solution which satisfies (\ref{chiral})
as well.  This sector preserves 1/4 supersymmetries.

We can say something else about the integrability condition (\ref{int'}) which
is independent of particular examples. Consider first the case $\sigma_1=
\sigma_2$.
We multiply Eq. (\ref{int'}) by  $F$. Using the identity  $-{1 \over 2}
\{\gamma_{ab},\gamma_{cd}\} = \epsilon_{abcd}\gamma_5 + \delta_{[ab][cd]}$ plus
the selfduality of $R^{ab}$ and chirality of $\epsilon_k$ one finds,
\begin{equation}
 R^{ab}\mbox{\tiny $\wedge$} R_{ab}\, \epsilon_k =
 F\mbox{\tiny $\wedge$} F\, \epsilon_k
\label{int'''}
\end{equation}
which becomes an identity on the selfdual background satisfying (\ref{RR=FF}).
Regarding the sector $\sigma_1=-\sigma_2$, we note that any selfdual linear
combination of Dirac
matrices $R^{ab} \gamma_{ab}$ has two zero eigenvalues. Thus (\ref{int'})
always have non-zero solutions.  This analysis indicates that all selfdual
solutions preserve some supersymmetry.

Note that the right hand side of (\ref{int}) contains a factor $1/N$. This
factor is present in all the following equations (\ref{int'}), (\ref{int''})
and (\ref{int'''}). Non-zero Killing spinors exist only for $1/N=1/2$, as
expected from (\ref{N=2}). In this sense, the Killing spinor equation provides
an independent derivation for the condition $N=2$.

To conclude, we have explored five-dimensional $SU(2,2|2)$ Chern-Simons
supergravity compactified to four dimensions, and show that exact, selfdual,
solutions exist. These solutions preserve some supersymmetry and we have
exhibited some explicit examples. More general solutions will be presented elsewhere
\cite{mbanados}.

The author would like to thank A. Chamblin, G. Gibbons, M. Henneaux, A. Ritz, R.
Troncoso and J. Zanelli for discussions and useful correspondence.  This work was
supported in part by FONDECYT (Chile)  grant \# 1000744 and DICYT (USACH) \#
04-0031BL.

\end{document}